\documentclass{icsm}
\usepackage{graphicx}% Include figure files
\usepackage{bm}% bold math
\usepackage{amsmath}
\usepackage{amsfonts}
\usepackage{amssymb}
\usepackage{amsthm}

\usepackage[dvips]{epsfig}

\slacs{.4ex}

\begin{document}
\title{Partition functions and graphs: A combinatorial approach}
\authori{A. I. Solomon$^{a,b}$, P. Bl{}asiak$^{b,c}$,}
\addressi{$^{a}$The Open University, Physics and Astronomy Department\\
Milton Keynes MK7 6AA, United Kingdom\\
$^{b}$Laboratoire de Physique Th\'eorique des Liquides,
Universit\'e Pierre et Marie Curie\\
Tour 24 -- 2e \'et., 4 Pl.Jussieu, F 75252 Paris Cedex 05, France\\
e-mail: a.i.solomon@open.ac.uk, blasiak@lptl.jussieu.fr
\\
$^{c}$H.Niewodnicza\'nski Institute of Nuclear Physics, Polish Academy of Sciences\\
ul. Eliasza-Radzikowskiego 152, PL 31342 Krakow, Poland\\}
\authorii{G. E. H. Duchamp,}
\addressii{Universit\'{e} de Rouen, LIFAR, F 76821 Mont-Saint Aignan Cedex, France\\
e-mail: gduchamp2@free.fr}
\authoriii{A. Horzela,}
\addressiii{H.Niewodnicza\'nski Institute of Nuclear Physics, Polish Academy of Sciences\\
ul. Eliasza-Radzikowskiego 152, PL 31342 Krakow, Poland\\
e-mail: andrzej.horzela@ifj.edu.pl}
\authoriv{K. A. Penson}
\addressiv{Laboratoire de Physique Th\'eorique des Liquides,
Universit\'e Pierre \& Marie Curie\\
Tour 24 -- 2e \'et., 4 Pl.Jussieu, F 75252 Paris Cedex 05, France\\
e-mail: penson@lptl.jussieu.fr}
\authorv{}    \addressv{}
\authorvi{}    \addressvi{}
\headauthor{A.I. Solomon et al.}
\headtitle{Partition functions and graphs: A combinatorial approach}
\lastevenhead{A.I. Solomon et al.: Partition functions and graphs:
A combinatorial approach}
\pacs{03.65.Fd, 05.30.Jp,}
\keywords{boson normal ordering, combinatorics}
\maketitle

\begin{abstract}
Although symmetry methods and analysis are a necessary ingredient
in every physicist's toolkit, rather less use has been made of
combinatorial methods. One exception is in the realm of
Statistical Physics, where the calculation of the partition
function, for example, is essentially a combinatorial problem. In
this talk we shall show that one approach is via the normal
ordering of the second quantized operators appearing in the
partition function. This in turn leads to a combinatorial
graphical description, giving essentially 'Feynman--type' graphs
associated with the theory. We illustrate this methodology by the
explicit calculation of two model examples, the free boson gas and
a superfluid boson model. We show how the calculation of partition
functions can be facilitated by knowledge of the combinatorics of
the boson normal ordering problem; this naturally gives rise to
the {\em Bell} numbers of combinatorics. The associated graphical
representation of these numbers gives a perturbation expansion in
terms  of a sequence of graphs analogous to zero--dimensional
Feynman diagrams.
\end{abstract}

\section{Introduction}

According to the axiomatic approach, it is possible to reconstruct
a quantum field theory from the vacuum expectation values. When we
{\em normally order} a set of operators so that all the
annihilation operators are to the right, then in such a vacuum
expectation value only the constant term remains.\footnote{Note
that this process of normally ordering, denoted by $\mathcal{N}$,
does {\em not} change the value of the operator. Thus
$\mathcal{N}(f(a,a^{\dagger}))=f(a,a^{\dagger})$.} Similarly, when
we normally order a string of boson operators $a$,$a ^{\dagger}$
satisfying $[a,a ^{\dagger}]=1$ , then the vacuum expectation
reduces to the constant term; and likewise we may recover the
expectation in the coherent state $|z\rangle$ by replacing $a$ by
the $c$--number $z$. Therefore the ability to normally order a
string of operators is a powerful calculational tool in physics.
It was recognized early on that the procedure of normally ordering
a string of bosons  leads to classical combinatorial numbers, the
Stirling and Bell numbers \cite{[1]}. Apart from the generation of
interesting combinatorial sequences, and their extensions
\cite{[2]}, the associated graph theory representations of these
numbers have analogies to the Feynman diagrams of a
zero--dimensional quantum field theory.

In this note we shall present a standard  graphical
representation of the Bell numbers which we show  is in some
sense {\em generic}. That is, we use these graphs to provide a
representation of a perturbation expansion for the partition
function of a free boson gas, and show that the same sequence of
graphs can be used to give a perturbation expansion for a general
partition function.

We illustrate this approach by applying it to a superfluid boson
system.

\section{Combinatorial sequences, generating functions and graphs}

\subsection{Some combinatorial sequences}

Sequences of numbers of combinatorial interest are common in
mathematics; for example, some sequences depending on one
parameter are
\begin{enumerate}
\item $\{n\}$  The integers.
\item $\{n!\}$ The number of ways of putting $n$ different objects into $n$ different
containers (leaving none empty).
\item $\{2^n\}$ The number of maps from an $n$--element set to $\{0,1\}$.
\item $\{B(n)\}$  The Bell numbers; the number of ways of putting $n$
different objects into $n$ identical containers (some may be left
empty).
\end{enumerate}
These are all familiar except perhaps the last, the Bell numbers.
These have the values
$$
B(n)=1,\,2,\,5,\,15,\,52,\,203,\,\ldots,\quad n=1,\,2,\,\ldots\,.
$$
and $B(n)\leq n!$.

We may also define sequences which depend on, say, two parameters,
for exam\-ple the 'choose' symbols
$^nC_k\equiv\dfrac{n!}{(n-k)!\;k!}$.  Related to the Bell numbers
are the {\em Stirling numbers of the second kind} $S(n,k)$, which
are defined as the number of ways of putting $n$ different objects
into $k$ identical containers, leaving none empty. From the
definition we have
\begin{equation}\label{sti}
B(n)=\sum_{k=1}^n S(n,k)\,.
\end{equation}

\subsection{Normal order}

Although somewhat unfamiliar to physicists, the Bell and Stirling
numbers are fundamental in quantum theory. This is because they
arise naturally in the {\em normal ordering problem}.  For
canonical bosons $[a,a^\dag]=1$ the  Stirling numbers of the
second kind $S(n,k)$ intervene through \cite{[3]}
\begin{eqnarray}
(a^\dag a)^n=\sum_{k=1}^nS(n,k) (a^\dag)^k a^k\,.
\end{eqnarray}
The corresponding Bell numbers $B(n)=\sum_{k=1}^n S(n,k)$ are
simply the expectation values
\begin{equation}\label{Bell}
 B(n)=\langle z|(a^{\dagger}a)^n|z\rangle_{z=1}
\end{equation}
taken in the coherent state defined by
\begin{equation}\label{cs}
 a|z\rangle=z|z\rangle
\end{equation}
for $z=1$. In fact, for physicists, these equations may be taken
as the {\em definitions} of the Stirling and Bell numbers.

\subsection{Generating functions}

Corresponding to a sequence of combinatorial numbers $\{a_n\}$ we
may define the {\em exponential generating function (egf)}
\begin{equation}\label{egf}
A(x)=\sum_{n=0}^\infty a_n\frac{x^n}{n!}\,.
\end{equation}
This provides a compact formula for the combinatorial sequence.
For example for the sequences above
\begin{enumerate}
\item $a_n=n$,\quad $A(x)=x\E^x$,
\item $a_n=n!$,\quad $A(x)=1/(1-x)$,
\item $a_n=2^n$,\quad $A(x)=\E^{2x}$,
\item $a_n=B(n)$,\quad $A(x)=\exp\left(\E^{x}-1\right)$.
\end{enumerate}
Except for the last, which we shall prove in what follows, these
exponential generating functions are immediate. Note that we
consider the expansion in a {\em formal} sense; we are not
concerned here about the convergence of the series defined by the
egf's. We may similarly define egf's corresponding to 2--parameter
sequences, thus:
\begin{equation}\label{egf1}
A(x,y)=\sum_{n=0}^\infty\biggl(\;\sum_{k=0}^\infty
a_{n,k}y^k\biggr) \frac{x^n}{n!}\,.
\end{equation}
For the 2--parameter sequences above we obtain:
\begin{enumerate}
\item $a_{n,k}\equiv ^n\!C_k$,\quad $A(x,y)=\exp((1+y)x)$,
\item $a_{n,k}\equiv S(n,k)$,\quad $A(x,y)=\exp\bigl(y(\E^{x}-1)\bigr)$.
\end{enumerate}
In the latter function, the coefficient of the $n$--th power of
$x$ is an $n$--degree polynomial in $y$, a Bell polynomial,
$B_n(y)$.

Therefore we have the generating function for the Bell numbers and
Bell polynomials
\begin{eqnarray}\label{bgf}
A_{B}(x)&=&\sum_{n=0}^\infty B(n)\,\frac{x^n}{n!}= \nonumber \\
&=&\exp\left(x+\frac{x^2}{2!}+\frac{x^3}{3!}+\ldots\right)
\end{eqnarray}
and
\begin{eqnarray}\label{bpgf}
A_{B}(x,y)&=&\sum_{n=0}^\infty\biggl(\;\sum_{k=0}^\infty
S(n,k)y^k\biggr)\frac{x^n}{n!}= \nonumber \\
&=&\sum_{n=0}^\infty B_n(y)\,\frac{x^n}{n!}= \nonumber \\
&=&\exp\bigl(y(\E^{x}-1)\bigr).
\end{eqnarray}

\subsection{Graphs}

We now give a graphical representation of the Bell numbers.
Consider labelled lines which emanate from a white dot, the
origin, and finish  on a black dot, the vertex. We shall allow
only one line from each white dot but impose no limit on the
number of lines ending on a black dot. Clearly this simulates the
definition of $S(n,k)$ and $B(n)$, with the white dots playing the
role of the distinguishable objects, whence the lines are
labelled, and the black dots that of the indistinguishable
containers. The identification of the graphs for 1,2 and 3 lines
is given  in the Figure 1.

\bfg[ht] \bc \resizebox{8.5cm}{!}{\includegraphics{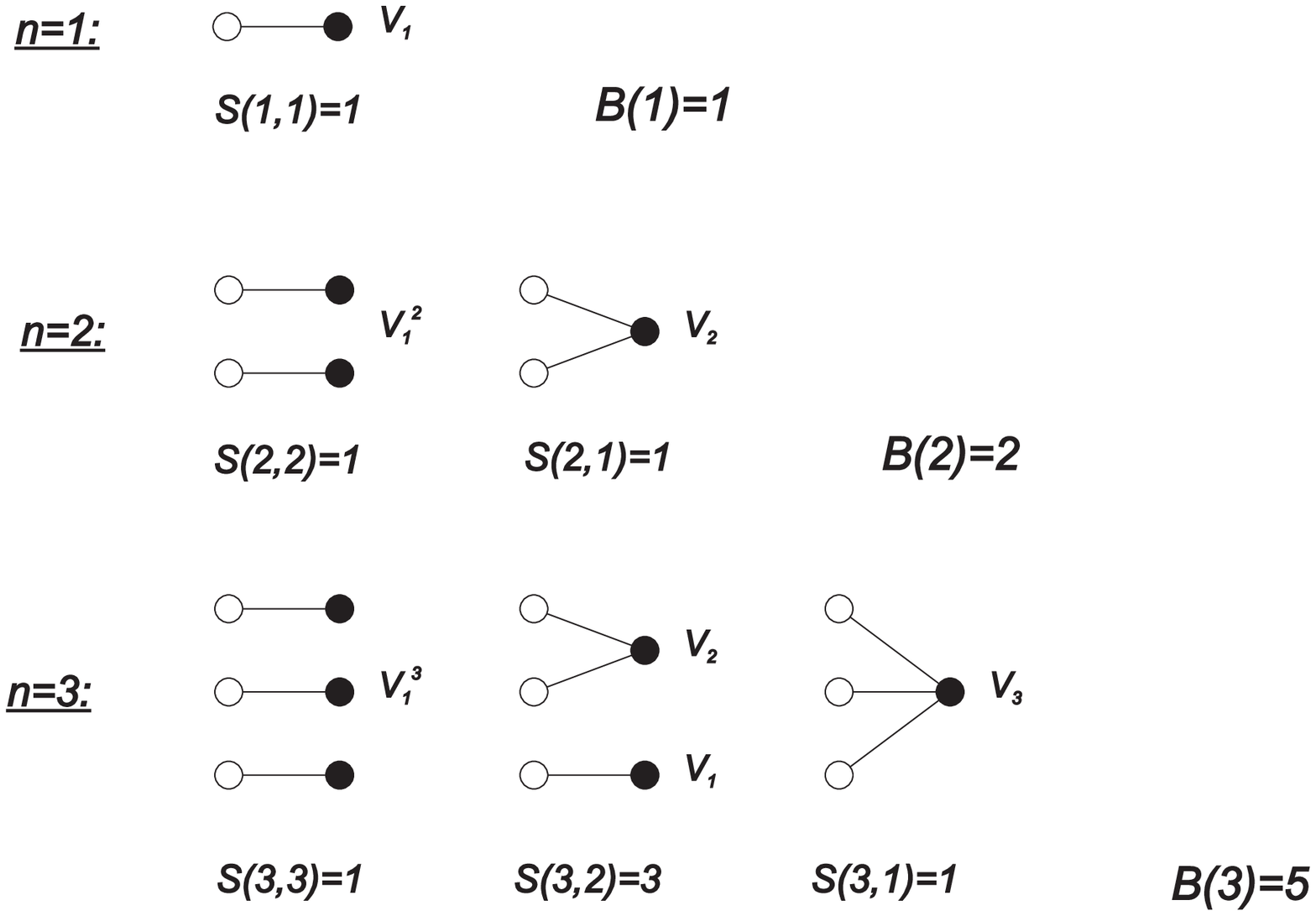}} \ec
\vspace{-7mm} \caption{Graphs for $B(n)$, $n=1$, 2, 3.}
\label{inter} \efg

We have concentrated on the Bell number sequence and its
associated graphs since, as we shall show, there is a sense in
which this sequence of graphs is {\em generic}. That is, we can
represent {\em any} combinatorial sequence by the same sequence of
graphs as in the Figure 1, with suitable vertex multipliers
(denoted by the $V$ terms in the same figure). One reason the
graphical representation is useful is that there exist some rather
powerful results which apply to graphs and their associated
generating functions. We give one such result now.

\subsection{Connected graph theorem}

This  states that if $C(x)=\sum_{n=1}^{\infty} c(n)x^n/n!$ is the
exponential generating function of {\em labelled connected}
graphs, {\em viz.} $c(n)$ counts the number of connected graphs of
order $n$, that is with $n$ lines, then
\begin{equation}
A(x) = \exp\bigl(C(x)\bigr)
\end{equation}
is the exponential generating function for {\em all}
graphs \cite{[4]}.

We may apply this very simply to the case of the $B(n)$  graphs
in  Figure 1. For each order $n$, the {\em connected} graphs
clearly consist of a single graph. Therefore for each $n$ we have
$c(n)=1$; whence, $C(x)=exp(x)-1.$ It follows that the generating
function for {\em all} the  graphs $A(x)$ is given by
\begin{equation}
A(x)=\exp\bigl(\exp(x) - 1\bigr)\,, \label{bell}
\end{equation}
which is therefore the generating function for the Bell numbers.

Given a general generating function
$C(x)=\sum_{n=1}^\infty{V_nx^{n}/n!}$ we may represent this by the
same sequence of graphs as in Figure 1, but now with vertex
factors ${V_n}= c(n)$, as shown in Figure 1.

This is of course a trivial observation, since any sequence of
graphs  each term of which consists of a single connected graph
will, with the appropriate multipliers $V_{n}$, give by definition
the exponential generating function of many combinatorial
sequences. This however shows the generic nature of the $B(n)$
sequence, as previously noted in \cite{[4a]}; and will prove of
value in our discussion of partition functions.

\section{Partition functions}

We first illustrate the use of normal ordering in the context of
{\em partition functions} by the most simple example. We define
as usual the partition function $Z$ associated with a Hamiltonian
$H$ by $Z\equiv {\rm Tr}\exp(-\beta H)$.

\subsection{Free boson gas and Bell polynomials}
\label{sub;ho}

We take as our example the Hamiltonian for the single--mode free
boson gas $H=\epsilon a^{\dagger}a$ (ignoring an additive
constant), $\epsilon>0$. The usual computation of the partition
function, exploiting the completeness property
$\sum_{n=0}^{\infty}|n\rangle\langle n|=I$, is immediate:
\begin{eqnarray}\label{pf1}
Z&=& {\rm Tr}\exp(-\beta \epsilon a^{\dagger}a)=\\
 &=& \sum_{n=0}^\infty{\langle n|\E^{-\beta \epsilon \hat{n}}|n\rangle}= \\
 &=& \sum _{n=0}^\infty{\E^{-\beta \epsilon n}}=\\
 &=&\left(1-\E^{-\beta \epsilon}\right)^{-1}\,.
\end{eqnarray}
However, we may, somewhat perversely in this simple case, use
{\em any} complete set to perform the trace. Choosing coherent
states, for which the completeness or {\em resolution of unity}
property is
\begin{equation}\label{cs1}
 \frac{1}{\pi}\int \D^{2}z |z\rangle\langle z|=I\equiv
 \int \D\mu(z)|z\rangle\langle z|
\end{equation}
the appropriate trace calculation is
\begin{eqnarray}\label{tr1}
Z&=&\frac{1}{\pi}\int \D^{2}z \langle z|\exp\bigl(-\beta \epsilon a^{\dagger}a\bigr)|z\rangle=\\
&=&\frac{1}{\pi}\int \D^{2}z \langle
z|:\exp\bigl(a^{\dagger}a(\E^{-\beta\epsilon}-1)\bigr):|z\rangle\,,
\end{eqnarray}
where we have used the following well-known relation
\cite{[5],[6]} for the {\em forgetful} normal ordering operator
$:\!f(a, a^{\dagger})\!:$ which means ``normally order the
creation and annihilation operators in $f$ {\em forgetting} the
commutation relation $[a,a^{\dagger}]=1$''\footnote{Of course,
this procedure may alter the value of the operator to which it is
applied.}:
\begin{equation}\label{no1}
\mathcal{N}\bigl(\exp(x
a^{\dagger}a)\bigr)=:\exp\bigl(a^{\dagger}a(\E^{x}-1)\bigr):.
\end{equation}
We therefore obtain, integrating over the angle variable $\theta$
and the radial variable $r=|z|$,
\begin{equation}
Z=\frac{1}{\pi}\int_0^{2\pi}\D\theta\int_0^\infty r\,\D r
\exp\left(r^2\bigl(\E^{-\beta \epsilon}-1\bigr)\right)\,,
\label{z1}
\end{equation}
which gives us $Z=\bigl(1-\E^{-\beta\epsilon}\bigr)^{-1}$ as
before.

We rewrite the above equation  to show the connection with our
previously--defined combinatorial numbers. Writing $y=r^2$ and
$x=-\beta \epsilon$, Eq.(\ref{z1}) becomes
\begin{equation}\label{z2}
Z=\int\limits_0^\infty \D y\exp\bigl(y(\E^{x}-1)\bigr)\,.
\end{equation}
This is an integral over  the classical generating function for
the Bell polynomials
\begin{equation}\label{bpgf1}
 \exp\left(y\bigl(\E^{x}-1\bigr)\right)=\sum_{n=0}^\infty B_n(y)\,\frac{x^n}{n!}\,,
\end{equation}
where $B_{n}(1)=B(n)$, as defined in Eq.(\ref{Bell}). This leads
to the combinatorial form for the partition function
\begin{equation}\label{z3}
Z=\int_0^\infty\D y\sum_{n=0}^\infty B_n(y)\,\frac{x^n}{n!}\,.
\end{equation}

Although Eq.(\ref{z3}) is remarkably simple in form, it is often
by no means a straightforward matter to evaluate the analogous
integral for other than the free boson system considered here.
Further, it is also clear that we may not interchange the integral
and the summation, as each individual $y$ integral diverges. We
shall therefore concentrate in  what follows on the {\em partition
function integrand} (PFI) $F(z)=\langle z|\exp(-\beta
H)|z\rangle$, whence $Z=\int F(z)\,\D\mu(z)$, to give a graphical
description of a perturbation approach. The function $F$ maps
coherent states $|z\rangle$ to (real) numbers, and is related to,
but not identical with, the previously-introduced {\em free energy
functional} \cite{[7]} associated with the Hamiltonian system.

\subsection{General partition functions}

We now apply this graphical approach to the general partition
function in second quantized form. With the usual definition for
the partition function
\begin{equation}\label{pf}
Z={\rm Tr}\exp(-\beta H)\,.
\end{equation}
In general the Hamiltonian is given by  $H=\epsilon
w(a,a^{\dagger})$, where $\epsilon$ is the energy scale, and $w$
is a string (= sum of products of positive powers) of boson
creation and annihilation operators. The partition function
integrand $F$ for which we seek to give a graphical expansion, is
\begin{equation}\label{pfz}
 Z(x)=\int{F(x,z)\,\D\mu(z)}\,,
\end{equation}
where
\begin{eqnarray}
F(x,z)&=&\langle z|\exp(xw)|z\rangle= \hskip20mm(x=-\beta \epsilon)\nonumber \\
&=&\sum_{n=0}^{\infty}\langle z|w^n|z\rangle\,\frac{x^n}{n!}=\nonumber \\
&=&\sum_{n=0}^{\infty}W_n(z)\,\frac{x^n}{n!}=\nonumber \\
&=&\exp\biggl(\;\sum_{n=1}^{\infty}V_n(z)\,\frac{x^n}{n!}\biggr),
\end{eqnarray}
with obvious definitions of $W_n$ and $V_n$. The sequences
$\left\{W_n\right\}$ and $\left\{V_n\right\}$ may each be
recursively obtained from the other \cite{[7a]}. This relates the
sequence of multipliers $\{V_n\}$ of Figure 1 to the Hamiltonian
of Eq.(\ref{pf}). The lower limit $1$ in the $V_n$ summation is a
consequence of the normalization of the coherent state
$|z\rangle$.

We conclude this discussion with an example.

\subsection{Superfluid boson system}

We may write the Hamiltonian for a superfluid boson system as
\cite{[8]}
\begin{equation}\label{sbs}
H=\epsilon\left\{\sfrac12\,c_1a^2+\sfrac12\,\bar{c}_1(a^{\dagger})^2+
c_2\bigl(a^{\dagger}a+\sfrac12\bigr)\right\}.
\end{equation}
As in the non-interacting boson gas example, this represents a
single mode of the system; the complete Hamiltonian is a direct
sum of non-interacting terms.  The constants $c_1$  and $c_2$ are
complex and real respectively, but otherwise arbitrary.  Noting
that the Hamiltonian Eq.(\ref{sbs}) is an element of the
non-compact Lie algebra $su(1,1)$ enables an expression for the
partition function integrand to be readily obtained (see the
Appendix). To illustrate the graphical series for this PFI, we
shall choose real arguments; that is  we put $z$ and  the constant
$c_1$ real in Eq.(\ref{pf8}) below to get
\begin{equation}\label{pf4}
 F(x,y)=\sqrt{\frac1{\mu}}\,\exp\left[y\left(y_1+\frac{1}{\mu}-1\right)\right],
\end{equation}
where $y=z^2$ and $y_1$, $\mu$ are functions of $x$. The PFI
$F(x,y)$ of Eq.(\ref{pf4}) generates an integer sequence for even
$c_1$, $c_2$ as may be easily verified using algebraic software.
For example, choosing $c_1=2$, $c_2=4$ we have
\begin{equation}\label{pf5}
F(x,y)=\exp\left(\frac{V_1(y)x}{1!}+\frac{V_2(y)x^2}{2!}+
\frac{V_3(y)x^3}{3!}+\ldots\right),
\end{equation}
with
\begin{equation}
\begin{array}{l}
V_1(y)=2+6y\,,\\
V_2(y)=2+36y\,,\\
V_3(y)=16+288y\,,\\
V_4(y)=144+3024y\,,\\
\ldots\,.
\end{array}
\end{equation}
leading to a corresponding Bell--type graphical expansion.

\section{Discussion}

The normal ordering of creation and annihilation operators leads
naturally to Stirling and Bell numbers. Therefore these
combinatorial numbers, and their graphical representations, are
ubiquitous in second--quantized quantum physics. In this talk we
emphasized this fact by application to a perturbative expansion
for the partition function of statistical physics. We did this
through the introduction of the {\em partition function integrand}
or PFI, a function which when integrated over the complete set of
coherent states gives the partition function proper. The
evaluation of the PFI depends on the normal ordering of the the
exponentiated Hamiltonian given in terms of  second--quantized
operators. In general this is a difficult if not intractable
problem, although approaches based on the {\em Product Theorem} of
combinatorial graph theory lead to some
simplification\footnote{See the talk by A. Horzela {\em et al} in
this conference.}. However, even with the straightforward methods
illustrated here, solutions are obtainable in many cases of
physical interest. In particular, we illustrated this approach in
the case of the free boson gas, and a superfluid boson model.  We
also gave a graphical representation for the perturbation
expansion of the PFI, an expansion in terms of $x=-\beta\epsilon$,
essentially energy scale times inverse temperature. The graphical
series is that of the Bell numbers, together with appropriate
multipliers $V_n(z)$, where $z$ is the coherent state parameter
which must be integrated over to obtain the partition function.
We showed how to determine these multipliers in terms of the
Hamiltonian. We conclude that there is a sense in which the Bell
graph series is generic for the evaluation of PFI's.

The operators of this theory are quantized, but have no explicit
dependence on space or time, which justifies considering the
theory as a zero--dimensional field theory, with the graphs as
being analogous to zero-dimensional Feynman diagrams.

\section*{Appendix: Partition function integrand for a superfluid boson system}

The partition function integrand $F(x,z)$ corresponding to the
hamiltonian Eq.(\ref{sbs})  is the expectation of the following
operator
\begin{eqnarray}
\exp(-\beta H)&=&\exp\left\{\sfrac12\,xc_1a^2+\sfrac12\,x\bar{c}_1(a^{\dagger})^2+
xc_2(a^{\dagger}a+\sfrac12)\right\}= \nonumber \\
&=&\exp\left\{\sfrac12\,x_1
a^2+\sfrac12\,\bar{x}_1(a^{\dagger})^2+x_2(a^{\dagger}a+\sfrac12)\right\}=
\qquad\qquad\left(\mathrm{for~}\ba{c}x_1=xc_1\,,\\x_2=xc_2\,,\ea\right)\nonumber \\
&=&\exp\{x_1 K^{-}+\bar{x}_1K^{+}+2x_2 K^{0}\}\,, \label{pf6}
\end{eqnarray}
where we have identified the generators $\{K^{+},K^{-},K^{0}\}$ of
$su(1,1)$ with the appropriate operators
$\bigl\{\frac12(a^{\dagger})^2,\frac12a^2,\frac{1}{2}(a^{\dagger}a+\frac12)\bigr\}$.
The normally ordered  form of  Eq.(\ref{pf6}) is
\begin{equation}\label{pf7}
\exp(-\beta H)=\E^{\bar{y}_1K^+}\E^{2y_2K^0}\E^{y_1K^-}\,.
\end{equation}
Since the equality of Eq.(\ref{pf6}) and Eq.(\ref{pf7}) is a
group--theoretical result, we may use the $2\times2$
representation
\begin{equation}\label{2x2}
{\slacs{1ex} \{K^{+},K^{-},K^{0}\}=\left\{\begin{pmatrix}
 0&1\\
 0&0 \end{pmatrix},
 \begin{pmatrix}
 0&0\\
 -1&0 \end{pmatrix},
 \begin{pmatrix}
 \frac{1}{2}&0\\
 0&\frac{1}{2} \end{pmatrix}
 \right\},}
\end{equation}
in which to implement the equality, thus evaluating $y_1$ and
$y_2$. The result is
$$
y_1=xc_1\sinh\left(\frac{\delta}{\mu\delta}\right),\quad
y_2=-\ln\mu\,,
$$
where
$$
\delta=x\sqrt{{c_2}^2-|c_1|^2}\,,\quad
\mu=\cosh\delta-xc_2\frac{\sinh\delta}{\delta}\,.
$$
We therefore have for the Partition Function Integrand
\begin{eqnarray}\label{pf8}
F(x,z)&=&\langle z|\exp\bigl\{\sfrac12\,x_1 a^2+\sfrac12\,\bar{x}_1(a^{\dagger})^2+
x_2 (a^{\dagger}a+\sfrac12)\bigr\}|z\rangle= \nonumber \\
&=&\bigl<z\bigr|\E^{\bar{y}_1 (a^{\dagger})^2/2}\E^{y_2(a^{\dagger}a+1/2)}
\E^{y_1 a^2/2}\bigl|z\bigr>= \nonumber \\
&=&\E^{\frac12\bar{y}_1\bar{z}^2}\E^{|z|^2(e^{y_2}-1)}\E^{y_1z^2/2}e^{\frac{1}{2}y_2}\,.
\end{eqnarray}

\bigskip\noindent
{\small P.B. wishes to thank the Polish Ministry of Scientific
Research and Information Technology for support under Grant no: 1
P03B 051 26}


\bbib{99}
\bibitem{[1]}
J. Katriel: Lett. Nuovo Cimento \textbf{10} (1974) 565.
\bibitem{[2]}
P. Blasiak, K.A. Penson, and A.I. Solomon: Ann. Comb. \textbf{7}
(2003) 127.
\bibitem{[3]}
J. Katriel: Phys. Lett. A. \textbf{273} (2000) 159.
\bibitem{[4]}
G.W. Ford and G.E. Uhlenbeck: Proc. Nat. Acad. \textbf{42} (1956)
122.
\bibitem{[4a]}
C.M. Bender, D.C. Brody, and B.K. Meister: J. Math. Phys.
\textbf{40} (1999) 3229.
\bibitem{[5]}
J.R. Klauder and E.C.G. Sudarshan: \textit{Fundamentals of Quantum
Optics}. Benjamin, New York, 1968.
\bibitem{[6]}
W.H. Louisell: \textit{Quantum Statistical Properties of
Radiation}. J. Wiley, New York, 1990.
\bibitem{[7]}
M. Rasetti: Int. J. Theor. Phys. \textbf{14} (1975) 1.
\bibitem{[7a]}
M. Pourahmadi: Amer. Math. Monthly \textbf{91} (1984) 303.
\bibitem{[8]}
A.I. Solomon: J. Math. Phys. \textbf{12} (1971) 390.
\ebib
\end{document}